\documentclass[prl,showpacs,superscriptaddress,amsmath,amssymb,twocolumn,aps]{revtex4}
\usepackage{graphicx}
\usepackage{dcolumn}
\usepackage{bm}

\begin{document}

\title{Experimental demonstration of the topological surface states protected by
the time-reversal symmetry}

\author{Tong Zhang}
\affiliation{Department of Physics, Tsinghua University, Beijing 100084, China}
\affiliation{Institute of Physics, Chinese Academy of Sciences,
Beijing 100080, China}
\author{Peng Cheng}
\affiliation{Department of Physics, Tsinghua University, Beijing
100084, China}
\author{Xi Chen}
\email{xc@mail.tsinghua.edu.cn}
\affiliation{Department of Physics, Tsinghua University, Beijing
100084, China}
\author{Jin-Feng Jia}
\affiliation{Department of Physics, Tsinghua University, Beijing 100084, China}
\author{Xucun Ma}
\affiliation{Institute of Physics, Chinese Academy of Sciences,
Beijing 100080, China}
\author{Ke He}
\affiliation{Institute of Physics, Chinese Academy of Sciences,
Beijing 100080, China}
\author{Lili Wang}
\affiliation{Institute of Physics, Chinese Academy of Sciences,
Beijing 100080, China}
\author{Haijun Zhang}
\affiliation{Institute of Physics, Chinese Academy of Sciences,
Beijing 100080, China}
\author{Xi Dai}
\affiliation{Institute of Physics, Chinese Academy of Sciences,
Beijing 100080, China}
\author{Zhong Fang}
\affiliation{Institute of Physics, Chinese Academy of Sciences,
Beijing 100080, China}
\author{Xincheng Xie}
\affiliation{Institute of Physics, Chinese Academy of Sciences,
Beijing 100080, China}
\author{Qi-Kun Xue}
\email{qkxue@mail.tsinghua.edu.cn} \affiliation{Department of
Physics, Tsinghua University, Beijing 100084, China}
\affiliation{Institute of Physics, Chinese Academy of Sciences,
Beijing 100080, China}

\date{\today}

\begin{abstract}
We report direct imaging of standing waves of the nontrivial surface states of
topological insulator Bi$_2$Te$_3$ by using a low temperature scanning tunneling
microscope. The interference fringes are caused by the scattering of the
topological states off Ag impurities and step edges on the Bi$_2$Te$_3$(111) surface.
By studying the voltage-dependent standing wave patterns, we determine the
energy dispersion $E(k)$, which confirms the Dirac cone structure of the
topological states. We further show that, very different from the conventional
surface states, the backscattering of the topological states by
nonmagnetic impurities is completely suppressed. The absence of backscattering is a spectacular manifestation of the time-reversal symmetry, which
offers a direct proof of the topological nature of the surface states.
\end{abstract}


\maketitle

The strong spin-orbital coupling in a certain class of materials gives rise to the
novel topological insulators in two~\cite{zhang06,molenkamp07} and three dimensions 
~\cite{kane07,moore07,zhang08a,kane08a,fang09} in the absence of
an external magnetic field. The topological states on the surfaces of three dimensional
(3D) materials have been studied recently in Bi$_{1-x}$Sb$_x$~\cite{kane08a,hasan08,hasan09a}, Bi$_2$Te$_3$ and Bi$_2$Se$_3$ 
~\cite{fang09,shen09,hasan09b,hasan09c,hasan09d}, which possess insulating gaps in the bulk and gapless states on surfaces. The
surface states of a 3D topological insulator comprise of an odd number of massless
Dirac cones and the crossing of two dispersion branches with opposite spins is fully
protected by the time-reversal-symmetry at the Dirac points. Such spin-helical states
are expected to bring forward exotic physics, such as magnetic monopole~\cite{zhang09a} and
Majorana fermions~\cite{kane08b,zhang09b}. To date, the experimental study of topological insulators
is predominantly limited to the determination of their band structure by angle-resolve
photoemission spectroscopy (ARPES) ~\cite{hasan08,hasan09a,shen09,hasan09b,hasan09c,hasan09d}. Distinct quantum phenomena
associated with the nontrivial topological electronic states still remain unexplored.
Particularly, there is no direct experimental evidence for the time reversal symmetry
that protects the topological property. Here, using the low temperature scanning
tunneling microscopy (STM) and spectroscopy (STS), we report the direct
observation of quantum interference caused by scattering of the 2D topologically
nontrivial surface states off impurities and surface steps. Our work strongly supports
the surface nature of the topological states, and provides a way to study the spinor
wave function of the topological state. More significantly, we find that the backscattering
of topological states by a nonmagnetic impurity is forbidden. This result
directly demonstrates that the surface states are indeed quantum-mechanically
protected by the time reversal symmetry.

The interference patterns in STM experiments~\cite{eigler93a,eigler93b,avouris93,heller03} result from the 2D surface
states perturbed by surface defects. A surface state is uniquely characterized by a 2D
Bloch wave vector $\vec{k}$ within the surface Brillouin zone (SBZ). During elastic
scattering, a defect scatters the incident wave with a wave vector $\vec{k}_i$ into
$\vec{k}_f =\vec{k}_i + \vec{q}$, with $\vec{k}_i$ and $\vec{k}_f$
being on the same constant-energy contour (CEC). The quantum
interference between the initial and final states results in a standing wave pattern
whose spatial period is given by $2\pi/q$ . When the STM images of a standing wave
are Fourier transformed~\cite{hoffman02}, the scattering wave vector $\vec{q}$ ($\hbar\vec{q}$
is the momentum
transfer) becomes directly visible in the reciprocal space.
In contrast, for bulk states, there will be continuous ranges of wave vectors on the
projected SBZ for a given energy. Usually, no distinct interference fringe can be
produced by bulk states and visualized by STM. In this sense, the standing wave is
surface-states-sensitive and particularly suitable for studying topological insulators.

Our experiments were conducted in an ultra-high vacuum low temperature (down
to 0.4 K) STM system equipped with molecular beam epitaxy (MBE) for film growth
(Unisoku). The stoichiometric Bi$_2$Te$_3$ film, a robust topological insulator, was
prepared on single crystal substrate Si(111) by MBE. Details of sample preparation
are described elsewhere~\cite{xue09}. Shown in Fig. 1(a) is a typical STM image of the Bi$_2$Te$_3$
film with a thickness of $\sim$100 nm. The atomically flat morphology of the film is
clearly observed. The three steps seen in Fig. 1(a) all have the height (0.94 nm) of a
quintuple layer. The steps are preferentially oriented along the three close-packing
([100], [110] and [010]) directions. The image with atomic resolution [Fig. 1(b)]
exhibits the two-dimensional hexagonal lattice structure of the Te-terminated (111)
surface of Bi$_2$Te$_3$. Our STM observation further reveals a small density of clovershaped
defects on the surface (see supporting material~\cite{supporting}). Similar to Bi$_2$Se$_3$~\cite{hasan09d,mahanti02,mahanti04}, these structures
can be assigned to the substitutional Bi defects at the Te sites by examining their
registration with respect to the $1\times1$-Te lattice in the topmost layer.

\begin{figure}[tbp]
        \includegraphics[width=3in]{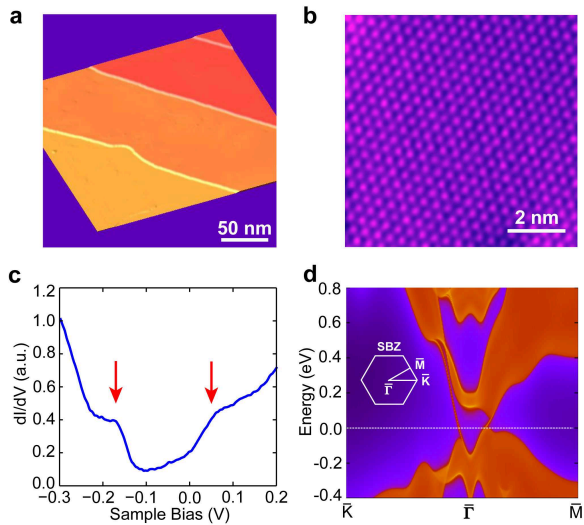}
        \caption{\label{fig1} (a) The STM topograph of the Bi$_2$Te$_3$(111) film. The imaged area, 250 nm by
250 nm, was scanned at a sample bias of 3 V and tunneling current of 50 pA. (b) The
atomic-resolution image. Tellurium atom (pink colored) spacing is about 4.3 \AA. The
image was scanned at a sample bias of -40 mV and tunneling current of 0.1 nA. (c)
$dI/dV$ spectrum taken on bare Bi$_2$Te$_3$(111) surface. The spectrum was measured with
setpoint V=0.3 V, I=0.1 nA. The arrows indicate the bottom of conduction band and
the top of valence band, respectively. (d) Calculated band structure of Bi$_2$Te$_3$(111)
along high-symmetry directions of SBZ (see the insert). The red regions indicate bulk
energy bands and the purple regions indicate bulk energy gaps. The surface states are
red lines around the $\bar{\Gamma}$ point. }
\end{figure}

The surface states of Bi$_2$Te$_3$ were investigated by STS and the first-principles
calculations~\cite{fang09}. The STS detects the differential tunneling conductance $dI/dV$
[Fig. 1(c)], which gives a measure of the local density of states (LDOS) of electrons at energy
eV. The two shoulders indicated by arrows in
Fig. 1(c) are the bottom of the bulk conduction band and the top of the bulk valence
band, respectively. The Fermi level (zero bias) is within the energy gap, indicating
that the film is an intrinsic bulk insulator~\cite{xue09}. The differential conductance in the
bulk insulating gap linearly depends on the bias and is attributed to the gapless
surface states. These features in STS are in good agreement with those obtained by
the first-principles calculations (see supporting material~\cite{supporting}). According to the calculations [Fig. 1(d)], the
topological states of Bi$_2$Te$_3$ form a single Dirac cone at the center ($\bar{\Gamma}$ point) of the
SBZ~\cite{shen09,hasan09b}, giving rise to a vanishing DOS in the vicinity of $k=0$. However, the
surface states around $\bar{\Gamma}$ point overlap in energy with the bulk valence band. For this
reason, the Dirac point is invisible in STS.

On the aforementioned surface, we deposited a small amount (0.01 ML) of Ag
atoms, which form trimmers on the surface, as
shown in Fig. 2(a) and more clearly in supporting material~\cite{supporting}. The atomically resolved STM image
(see supporting material~\cite{supporting}) reveals that the Ag atom in a trimmer adsorbs on the top site of a surface Te
atom~\cite{geometry}. This situation is schematically shown in Fig. 2(b). In addition, the Fermi
level shifts upwards in energy by 20$\sim$30 meV after Ag deposition, suggesting electron
transfer from the Ag atoms to the substrate (see supporting material~\cite{supporting}). The $dI/dV$ mapping was then
carried out in a region containing Ag trimmers. At each data point, the feedback was
turned off and the bias modulation was turned on to record $dI/dV$. This procedure
resulted in a series of spatial mapping of LDOS at various bias voltages. 
 
 \begin{figure*}[t]
        \includegraphics[width=6in]{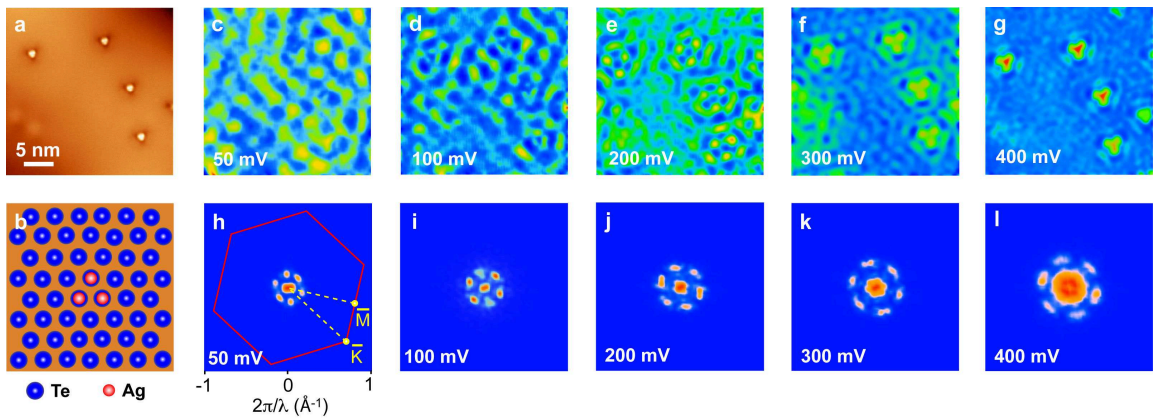}
        \caption{\label{fig2} Standing waves induced by Ag trimmers on Bi$_2$Te$_3$(111) surface. (a) STM
image (28 nm by 28 nm) of a region with four Ag trimmers adsorbed on Bi$_2$Te$_3$(111)
surface. (b) The adsorption geometry of Ag trimmer. (c) to (g) The $dI/dV$ maps of the
same area as (a) at various sample bias voltages. Imaging conditions: I=0.1 nA. Each map has 128 by 128 pixels and
took two hours to complete. The interference fringes are evident in the images. The
green and red regions indicate modulations with high intensity. (h) to (k) The FFT
power spectra of the $dI/dV$ maps in (c) to (g). The SBZ in (h) is superimposed on
the power spectra to indicate the directions in $\vec{q}$-space. The resolution of FFT, which
is $2\pi/28$ nm$^{-1}$, is determined by the size of the STM image. }
\end{figure*}

Figures 2(c) to 2(g) summarize the $dI/dV$ maps for bias voltages ranging from 50
mV to 400 mV from the area shown in Fig. 2(a). The first striking aspect of these
images is the existence of standing wave~\cite{eigler93a,eigler93b,avouris93} in the vicinity of the Ag trimmers.
The spatial modulation of LDOS by an Ag trimmer forms a hexagonal pattern, whose
edges are perpendicular to the $\bar{\Gamma}-\bar{\text{M}}$ directions in SBZ. This situation is more clearly
resolved at large bias voltages [Figs. 2(f) and 2(g)]. As expected, the interference pattern is
anisotropic as a result of the hexagram CEC~\cite{shen09}. The spatial period of the standing
wave scales inversely with the bias voltage and is determined by the momentum
transfer during scattering at a given energy. Below 50 meV, the fringes become
obscured. It results from a combination of two effects: (i) The wavelength increases
rapidly as the bias voltage approaches the Dirac point, where $k=0$. (ii) At low energy,
more topological surface states with different wavelengths are involved in the
formation of standing wave as indicated by the first-principles calculations [Fig. 1(d)].
The superposition of waves with various wavelengths smears out the interference
fringes. With increasing bias, especially after the surface states in the $\bar{\Gamma}-\bar{\text{M}}$ direction
merge into the bulk conduction band at $\sim$0.2 eV above the Dirac point according to
calculation [Fig. 1(d)], the contribution of states in the $\bar{\Gamma}-\bar{\text{M}}$ direction vanishes and
the states in the vicinity of $\bar{\Gamma}-\bar{\text{K}}$ direction gradually gain more weight, leading to
more distinct interference patterns. After the surface states in the $\bar{\Gamma}-\bar{\text{K}}$ direction
merge into the bulk conduction band at $\sim$0.6 eV above the Dirac point [Fig. 1(d)], the
standing waves fade out again.

To quantify the standing waves and obtain the scattering wave vectors, we
performed Fourier transformation of the $dI/dV$ maps into the $\vec{q}$-space [Figs. 2(h) to 2(l)].
One important feature in the power spectra can be immediately discerned by
comparing the six-fold symmetric pattern in the $\vec{q}$-space with SBZ (the red hexagon
in Fig. 2(h)): the regions with high intensity are always oriented toward the $\bar{\Gamma}-\bar{\text{M}}$
directions, while the intensity in the $\bar{\Gamma}-\bar{\text{K}}$ directions vanishes (see supporting material~\cite{supporting}). Such
phenomena can be understood by exploring possible scattering processes on the CEC
in the reciprocal space [Fig. 3(a)]. Generally, the $\vec{k}_i$ and $\vec{k}_f$
pairs with high joint DOS should dominate the quantum interference. For energies at which the interference
fringes are prominent, the regions on CEC with high DOS are primarily centered
around the $\bar{\Gamma}-\bar{\text{K}}$ directions~\cite{shen09}. Therefore, three scattering wave vectors, labeled $\vec{q}_1$,
$\vec{q}_2$ and $\vec{q}_3$, might be expected to appear in the power spectra. Among them, however,
only $\vec{q}_2$ is along the $\bar{\Gamma}-\bar{\text{M}}$ directions and can generate the observed standing waves.
Both $\vec{q}_1$ and $\vec{q}_3$ are invisible in the power spectra. 
There is a simple argument to account for the
disappearance of $\vec{q}_1$: the time-reversal invariance. The topological states $|\vec{k}\uparrow\rangle$
and $|-\vec{k}\downarrow\rangle$ are related by the time-reversal transformation: $|-\vec{k}\downarrow\rangle=\mathcal{T}|\vec{k}\uparrow\rangle$, 
where $\mathcal{T}$ is the time-reversal operator. It is straightforward to show that
$\langle-\vec{k}\downarrow|U|\vec{k}\uparrow\rangle=-\langle\vec{k}\uparrow|U|-\vec{k}\downarrow\rangle^*=
-\langle-\vec{k}\downarrow|U|\vec{k}\uparrow\rangle=0$ for fermions, where $U$ is a
time-reversal invariant operator and represents the impurity potential of the nonmagnetic
Ag impurity. Therefore, the backscattering between $\vec{k}$ and $-\vec{k}$ is
quantum-mechanically prohibited. 
Most of the observed features in the interference pattern, including the extinction of wave vector $\vec{q}_3$, have
been recently well explained by a full theoretical treatment~\cite{wu09a} based on the T-matrix approach for multiband systems~\cite{wu09b}. 
In addition to the existence of standing waves, 
%
the absence of backscattering
represents the second and most striking aspect of our experiment, which
makes the topological standing waves more extraordinary as compared to the
conventional surface states on metal samples~\cite{eigler93a,eigler93b,avouris93,heller03}. 
 
\begin{figure}[h]
        \includegraphics[width=3in]{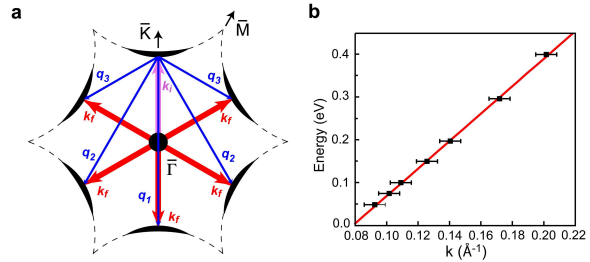}
        \caption{\label{fig3} (a) The scattering geometry. The CEC is in the shape of a hexagram. The
dominant scattering wave vectors connect two points in $\bar{\Gamma}-\bar{\text{K}}$ directions on CEC.
$\vec{k}_i$ (pink arrow) and $\vec{k}_f$ (red arrows) denote the wave vectors of incident and scattered
states. $\vec{q_1}$, $\vec{q_2}$ and $\vec{q_3}$ (blue arrows) are three possible scattering wave vectors. (b)
Energy dispersion as a function of $k$ in the $\bar{\Gamma}-\bar{\text{K}}$ direction. The data (black squares)
are derived from FFT in Fig. 2. The red line shows a linear fit to the data with $v_F=4.8\times10^5$ m/s. 
The error bars represent the resolution of FFT (see the caption of Fig. 2). }
\end{figure}

 We can obtain the dispersion of the massless Dirac fermions in the $\bar{\Gamma}-\bar{\text{K}}$
direction using the interference patterns and their Fourier transforms. For
$\vec{q}_2$, the scattering geometry determines $q_2 =\sqrt{3}k$ [see Fig. 3(a)], where $k$ is the wave vector in
the $\bar{\Gamma}-\bar{\text{K}}$ direction at a given energy. The resulting $k$ values vary linearly with
energy [Fig. 3(b)]. The slope of the linear fitting provides a measurement of the Dirac
fermion velocity $v_F$ , which is $4.8\times10^5$ m/s. In addition, the energy of the Dirac point
is estimated to be –0.25 eV by the intercept of the dispersion with the energy axis.
These observations are in agreement with the results from the first-principles
calculation and the ARPES data~\cite{fang09,xue09}. More importantly, the unoccupied states,
which are inaccessible to ARPES, can be probed by the standing waves with STM.

Interference fringes are also found at the step edges on the surface~\cite{kapitulnik09} [Figs. 4(a) to 4(h)].
Similar to the case of Ag trimmers, the standing waves produced by steps are
predominantly propagating along the $\bar{\Gamma}-\bar{\text{M}}$ direction. The fringes are clearly visible
even at the negative bias voltages probably owing to the stronger scattering potential
compared to that of the Ag trimmers. The dispersion curve deduced from these
patterns again shows a linear relation between the scattering wave vector and the
energy [Fig. 4(i)]. Using the slope of the linear fitting together with the same scattering
geometry as that for the Ag trimmer, the Fermi velocity is found to be $4.8\times10^5$ m/s,
the same as that obtained from the standing waves caused by the Ag impurities.
 
 \begin{figure}[h]
        \includegraphics[width=3in]{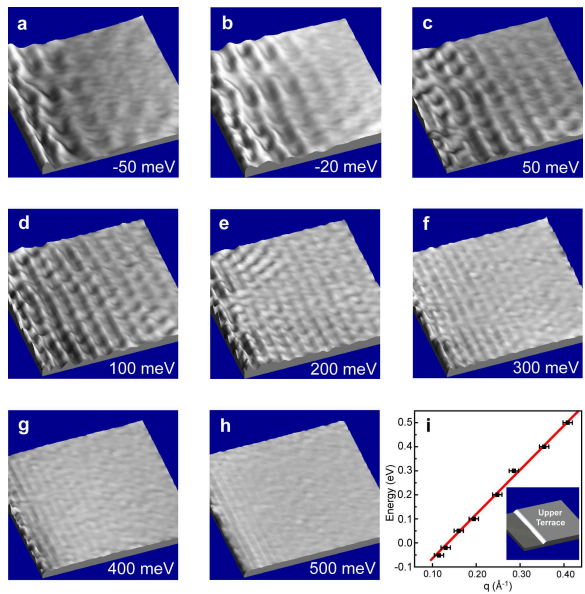}
        \caption{\label{fig4}Standing waves on the upper terrace by a step edge ((a) to (h)). All the images are
$dI/dV$ maps at various bias voltages of an area of 35 nm by 35 nm. Imaging conditions:
I=0.1 nA. (i) Energy dispersion deduced from the standing waves at the step edge.
The dispersion is a function of the scattering wave vector $q$. The inserted STM image
shows the step that produces the standing waves in (a) to (h).      
        }
\end{figure}
 
 The existence of standing wave strongly supports the surface nature of topological
states. An important issue that immediately arises is whether the topological states
respond differently to the magnetic and the nonmagnetic impurities. Theoretically, it
was pointed out~\cite{zhang08a,zhang09c,zhang08b} that a time-reversal breaking perturbation, such as
magnetic impurities, can induce scattering between the states $|\vec{k}\uparrow\rangle$
and $-|\vec{k}\downarrow\rangle$ and
open up a local energy gap at the Dirac point. It remains an open question to observe
the distinct signature of time-reversal breaking in topological insulators.

We thank S.-C. Zhang, X.-L. Qi, Y. Ran and S.-Q. Shen for
valuable discussions. The work is supported by NSFC and the National Basic
Research Program of China. The STM topographic images were processed using
WSxM (www.nanotec.es).

Note added. At the completion of this manuscript for submission, we became
aware of related work by P. Rousan {\it et al.}~\cite{yazdani09}. 
The authors reported STM study of scattering from disorder in BiSb alloy.


\begin{references}
\bibitem{zhang06}
B. A. Bernevig, T. L. Hughes, and S.-C. Zhang, Science {\bf 314}, 1757 (2006).
\bibitem{molenkamp07}
M. K\"{o}nig {\it et al.}, Science {\bf 318}, 766 (2007).
\bibitem{kane07}
L. Fu, C. L. Kane, and E. J. Mele, Phys. Rev. Lett. {\bf 98}, 106803 (2007).
\bibitem{moore07}
J. E. Moore and L. Balents, Phys. Rev. B {\bf 75}, 121306 (2007).
\bibitem{zhang08a}
X.-L. Qi, T. L. Hughes, and S.-C. Zhang, Phys. Rev. B {\bf 78}, 195424 (2008).
\bibitem{kane08a}
J. C. Y. Teo, L. Fu, and C. L. Kane, Phys. Rev. B {\bf 78}, 045426 (2008).
\bibitem{fang09}
H. J. Zhang {\it et al.}, Nature Phys. {\bf 5}, 438 (2009).
\bibitem{hasan08}
D. Hsieh {\it et al.}, Nature {\bf 452}, 970 (2008).
\bibitem{hasan09a}
D. Hsieh {\it et al.}, Science {\bf 323}, 919 (2009).
\bibitem{shen09}
Y. L. Chen {\it et al.}, Science {\bf 325}, 178 (2009).
\bibitem{hasan09b}
D. Hsieh {\it et al.}, Nature {\bf 460}, 1101 (2009).
\bibitem{hasan09c}
Y. Xia {\it et al.}, Nature Phys. {\bf 5}, 398 (2009).
\bibitem{hasan09d}
Y. S. Hor {\it et al.}, Phys. Rev. B {\bf 79}, 195208 (2009).
\bibitem{zhang09a}
X.-L. Qi, R. Li, J. Zang, and S.-C. Zhang, Science {\bf 323}, 1184 (2009).
\bibitem{kane08b}
L. Fu and C. L. Kane, Phys. Rev. Lett. {\bf 100}, 096407 (2008).
\bibitem{zhang09b}
X.-L. Qi, T. L. Hughes, S. Raghu, and S.-C. Zhang, Phys. Rev. Lett. {\bf 102}, 187001
(2009).
\bibitem{eigler93a}
M. F. Crommie, C. P. Lutz, and D. M. Eigler, Nature {\bf 363}, 524 (1993).
\bibitem{eigler93b}
M. F. Crommie, C. P. Lutz, and D. M. Eigler, Science {\bf 262}, 218 (1993).
\bibitem{avouris93}
Y. Hasegawa and Ph. Avouris, Phys. Rev. Lett. {\bf 71}, 1071 (1993).
\bibitem{heller03}
G. A. Fiete and E. J. Heller, Rev. Mod. Phys. {\bf 75}, 933 (2003).
\bibitem{hoffman02}
J. E. Hoffman {\it et al.}, Science {\bf 297}, 1148 (2002).
\bibitem{xue09}
Y. Y. Li {\it et al.} (unpublished).
\bibitem{supporting}
See EPAPS for additional materials about the experimental results.
\bibitem{mahanti02}
S. Urazhdin {\it et al.}, Phys. Rev. B {\bf 66}, 161306 (2002).
\bibitem{mahanti04}
S. Urazhdin {\it et al.}, Phys. Rev. B {\bf 69}, 085313 (2004).
\bibitem{geometry}
The other candidate model of the Ag trimmer is that the Ag atom substitutes a
topmost layer Te atom. In this case, the formation of Ag trimmers is kinetically
more difficult. Although the exact structure of Ag trimmers does not affect the
main conclusion here, it remains an interesting subject for further study, for
example, by first-principles calculation.
\bibitem{wu09a}
W.-C. Lee, C. J. Wu, D. P. Arovas, and S.-C. Zhang, arXiv:0910.1668 (2009).
\bibitem{wu09b}
W.-C. Lee and C. J. Wu, arXiv:0906.1973 (2009).
\bibitem{kapitulnik09}
Z. Alpichshev {\it et al.}, arXiv:0908.0371 (2009).
\bibitem{zhang09c}
Q. Liu, C.-X. Liu, C. Xu, X.-L. Qi, and S.-C. Zhang, Phys. Rev. Lett. {\bf 102}, 156603
(2009).
\bibitem{zhang08b}
X.-L. Qi, T. L. Hughes, and S.-C. Zhang, Nature Phys. {\bf 4}, 273 (2008).
\bibitem{yazdani09}
P. Rousan {\it et al.}, Nature {\bf 460}, 1106 (2009).

\end{references}
\end{document}